\documentclass{elsart}
\usepackage{graphicx}
\usepackage{amssymb}

\begin{document}

\begin{frontmatter}

\title{The network of concepts in written texts }
\center{
\author{Silvia M. G. Caldeira$^1$, Thierry C. Petit Lob\~{a}o$^2$,
R. F. S. Andrade$^1$,}
\author{ Alexis Neme$^3$ \and J. G. V. Miranda$^1$}}

\address{$^1$ Instituto de F\'isica, Universidade Federal da Bahia  \\
  Campus Universit\'ario da Federa\c c\~ao, 40210-340 - Salvador - BA, Brazil.\\
  $^2$ Instituto de Matem\'atica, Universidade Federal da Bahia  \\
  Campus Universit\'ario da Federa\c c\~ao, 40210-340 - Salvador - BA, Brazil.\\
  $^3$ Institut d'eletronique et d'informatique Gaspard-Monge, IGMFrance.}



\begin{abstract}
Complex network theory is used to investigate the structure of
meaningful concepts in written texts of individual authors.
Networks have been constructed after a two phase filtering, where
words with less meaning contents are eliminated, and all remaining
words are set to their canonical form, without any number, gender
or time flexion. Each sentence in the text is added to the network
as a clique. A large number of written texts have been
scrutinized, and its found that texts have small-world as well as
scale-free structures. The growth process of these networks has
also been investigated, and a universal evolution of network
quantifiers have been found among the set of texts written by
distinct authors. Further analyzes, based on shuffling procedures
taken either on the texts or on the constructed networks, provide
hints on the role played by the word frequency and sentence length
distributions to the network structure. Since the meaningful words
are related to concepts in the author's mind, results for text
networks may uncover patterns in communication and language
processes that occur in the mind.

\end{abstract}
\end{frontmatter}

\section{Introduction}

Concepts of complex networks have proven to be powerful tools in
the analysis of complex systems \cite{r1,r2,r3,r4,r5}. They have
been applied to modelling purposes as well as to search for
properties that naturally emerge in actual systems due to their
large-scale structure. Unlike random graphs, complex networks
reveals ordering principles related to their topological
structure. This way, if complex systems are mapped onto networks,
it is possible to use their conceptual framework to identify and
even to explain features that seem to have universal character.

Several complex networks have been proposed in the scientific
literature associated with real systems: the biological food
web\cite{r6}, technological communication networks as the
Internet, information networks as the World Wide Web\cite{r7},
social networks defined by friendship relations among individuals,
etc. \cite{r8}.

Word networks have been used to address complex aspects of human
language. In such studies, words are connected according either to
semantic associations \cite{r11,r12,r13} or even by nearness in
the text \cite{r14,r15,r16}, i.e., based on what is commonly
called word window with a fixed number of words. Those works
intend to establish the structure of a given language as a whole.
Because of this, they deal with a huge amount of texts,
independently of their authors, what is called corpora.

Language is obviously a product of brain activity and, to this
regard, we would like to call the attention to works revealing the
presence of neuronal networks in the brain by direct physiological
measurements.  In a recent work\cite{r17}, functional magnetic
resonance of human brains has established networks whose vertices
are regions of the cerebral cortex that are activated by external
stimuli according to a temporal correlation of activity.

In this work, we investigate the relations between the concepts in
individual written texts, by using them as starting point to
construct significant networks. Projecting both the concepts
present in the text, as well as the way they are related among
them, onto a network gives the opportunity to use the tools and
concepts developed within the network framework to characterize,
in a quantitative way, how the concepts in a written text appear,
how ordered and connected they are, how close to each other they
are within the text, and so on.

\section{Network concepts and text network construction}

An undirected network is defined by a set of elements, called
vertices or nodes, represented graphically by points, some of
which are joined together by an edge, represented by a line
between these points. The topological structure of many networks
displays a large-scale organization, with some typical properties
like: highly sparse connectivity, small minimal paths between
vertices and high local clustering,

To analyze a network, a number of indices (or quantifiers) have
been proposed, which allow for a quantification of many of the
properties quoted above. In this work we will use the most
important of them: number of vertices $N$, number of edges $M$,
average connectivity $<k>$, average minimal path length $\ell$,
diameter $L$, degree distribution $p(k)$ , and the Watts and
Strogatz  clustering coefficient $C$ (for a thorough discussion of
these indices see, e.g., \cite{r5}). If the node degree
distributions follow power laws, $p(k)\sim k^{-\gamma}$, the
exponent $\gamma$ becomes an important index for the network
characterization.

Two important network scenarios, characterized respectively by the
small-world \cite{r9} and scale-free structures \cite{r10}, have
been identified to be present in several complex networks,
including those related to the subject of this work, namely the
analysis of corpora and of brain activity. The first one is
related to the concomitant presence of small average value for
$\ell$ and high $C$, while the second is related to a power law
distribution of node degree.

To map the texts onto networks that reflect the structure of
meaning concepts within it, we have preserved only the words with
an intrinsic meaning, eliminating words which have a merely
grammatical function, as to arrange the syntactical structure of
sentences in the text (articles, pronouns, prepositions,
conjunctions, abbreviations, and interjections). Afterwards, we
have reduced the remaining words to their canonical forms, i.e.,
we have disregarded all inflections as plural forms, gender
variations, and verb forms. Such procedures are common in studies
on language\cite{r18}. In order to perform a computer
implementation, we have used some routines, dictionaries, and
grammatical rules from UNITEX package\cite{r19}. Unknown words to
the program were preserved in their original forms. After this
filtering treatment, we have constructed a network for each
individual text, where each distinct word corresponds to a single
vertex. Since we consider the sentence as the smaller significant
unit of the discourse, two words are connected if they belong to
the same sentence. Thus, sentences are incorporated into the
network as a complete subgraph, that is, a clique of mutually
connected words. Sentences with common words are connected by the
shared words. The method we propose offers a natural way to
analyze the growth process of the network evolution. We have
investigated both the network behavior in this step by step
evolving process, what we call dynamical analysis, as well as the
behavior of the entire network in its final form, corresponding to
the entire text, called the static analysis. Both methods reveal
important properties and they shall be discussed further on. In
Figure 1, we show an example of a text filter process together
with the evolving process of network construction for a well known
jingle.

\begin{figure}
\begin{center}
\includegraphics*[width=5cm,angle=270]{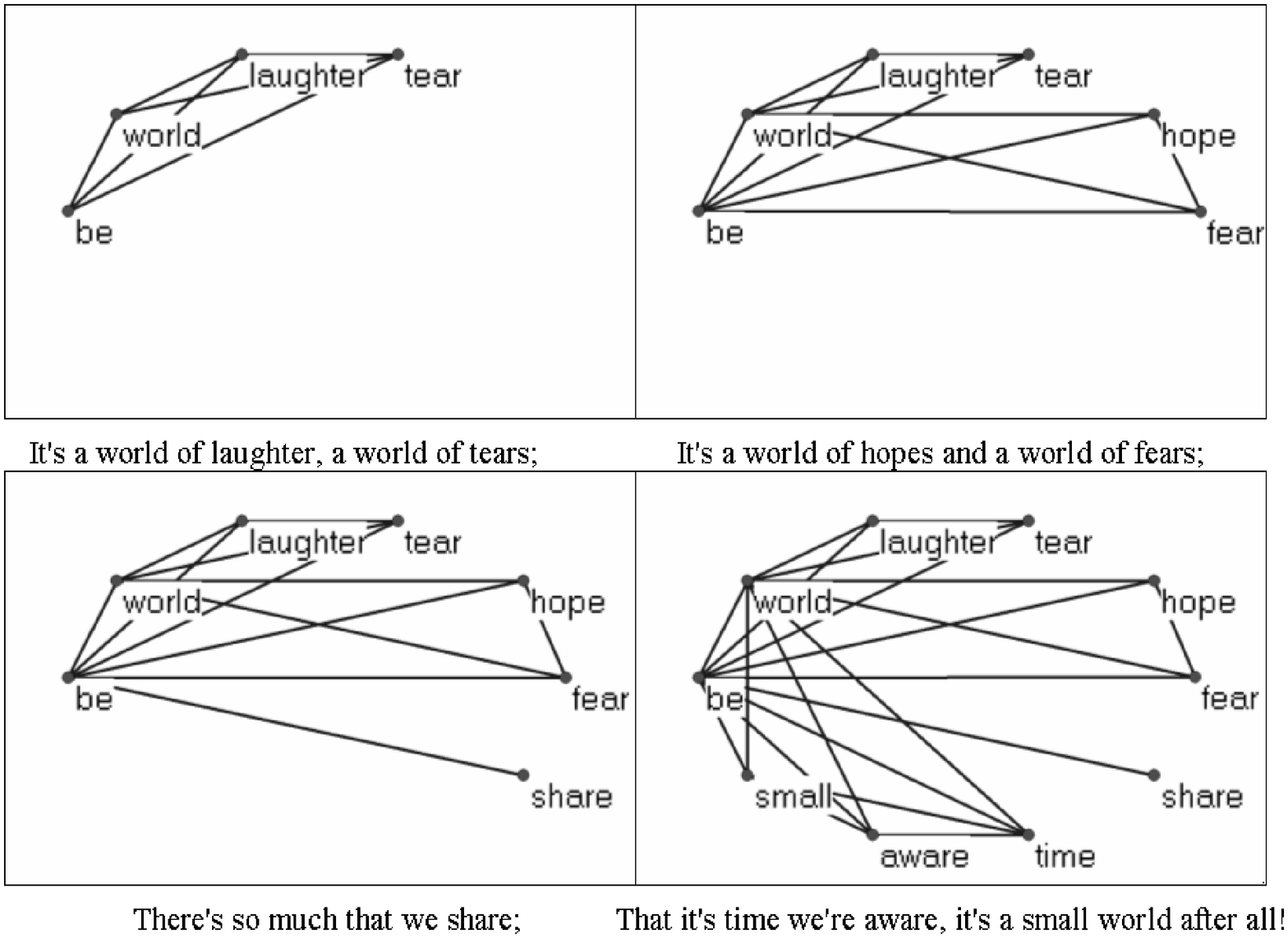}
\end{center}
\caption{The filtering treatment and the growth process of the
network for a small poem.}
\end{figure}

In order to guarantee an uniform and comprehensive sample, we have
chosen a collection of 312 texts \cite{roda1}, which can be cast
into different classes as: genre (59\% technical, 41\% literary),
language (53\% in Portuguese, 47\% in English), gender (72\% male
authors, 28\% female authors). Finally we have also classified the
texts according to their size (55\% with less than 1,000
sentences, 45\% with more than 1,000). The smaller text has 169
and the larger, 276,425 words; in the average, the texts have
32,691 words.

\section{Results}

The texts were individually analyzed, based on the evaluation of
the indices quoted in the previous section: $N$, $M$,  $<k>$,
$\ell$, $C$, $L$, and degree distribution $p(k)$, which we have
found to obey a power law  $p(k)\sim k^{-\gamma}$. We have also
followed how the number and size of independent clusters varies in
the growth process. Average values of such indices based on the
entire texts of the whole sample have been computed and are shown
in Table I. The analyzed networks are very sparse, with average
connection index $2M/[N(N-1)]<0.1$ for the networks with
$N\geq100$ .

We have verified that the degree distributions, for all analyzed
texts, exhibit an early maximum around $k=10$, which is followed
by crossover to a long right-skewed tail, approximating a power
law distribution with average exponent $\langle \gamma \rangle =
1.6 \pm 0.2$. We have found no evidence of correlation between the
value of $\gamma$ and the length of the analyzed text. These
features are illustrated in Figure 2, where we show $p(k)\times k$
for a very large text and the smallest one we analyzed.

\begin{figure}
\begin{center}
\includegraphics*[width=4cm,angle=270]{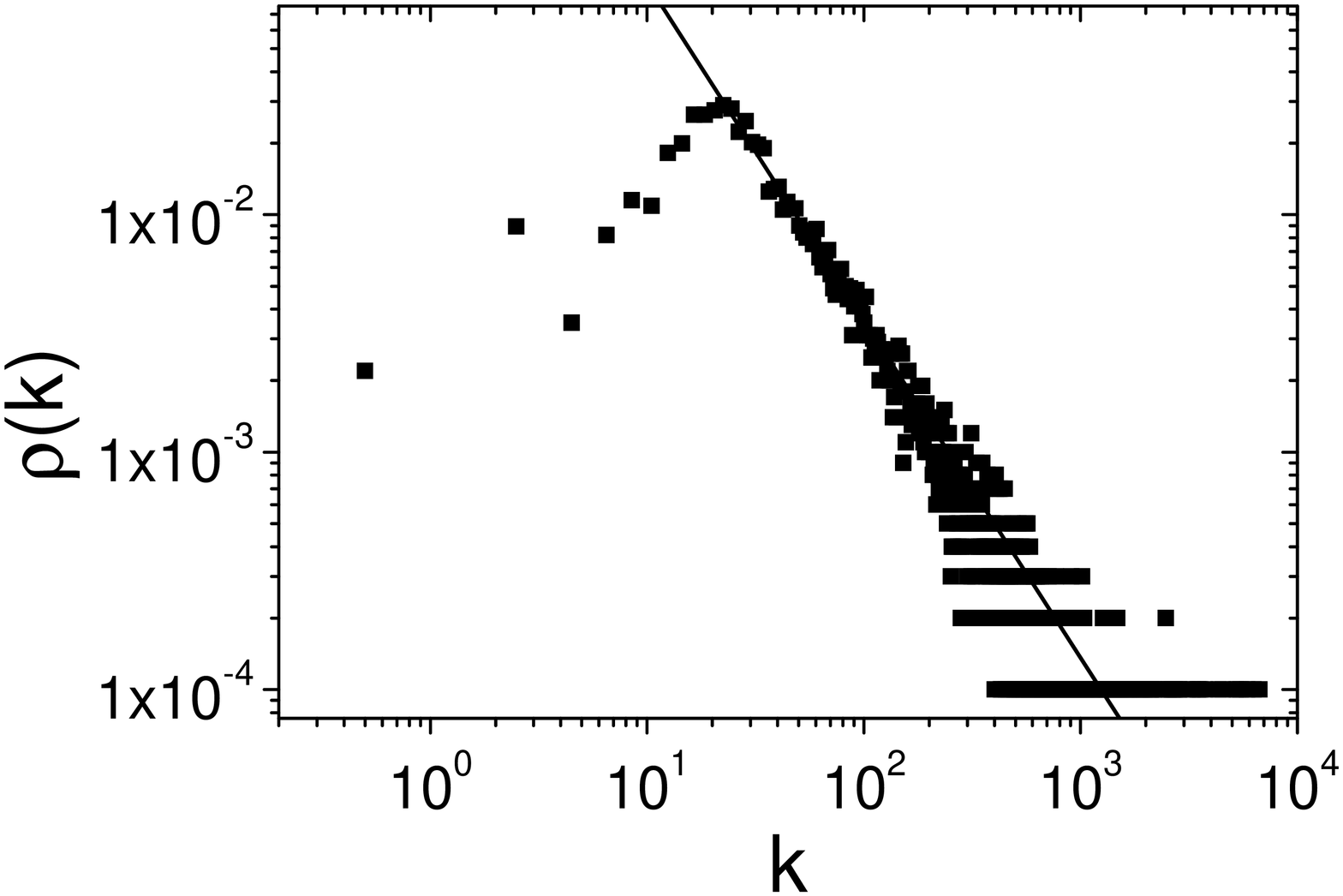}
\includegraphics*[width=4cm,angle=270]{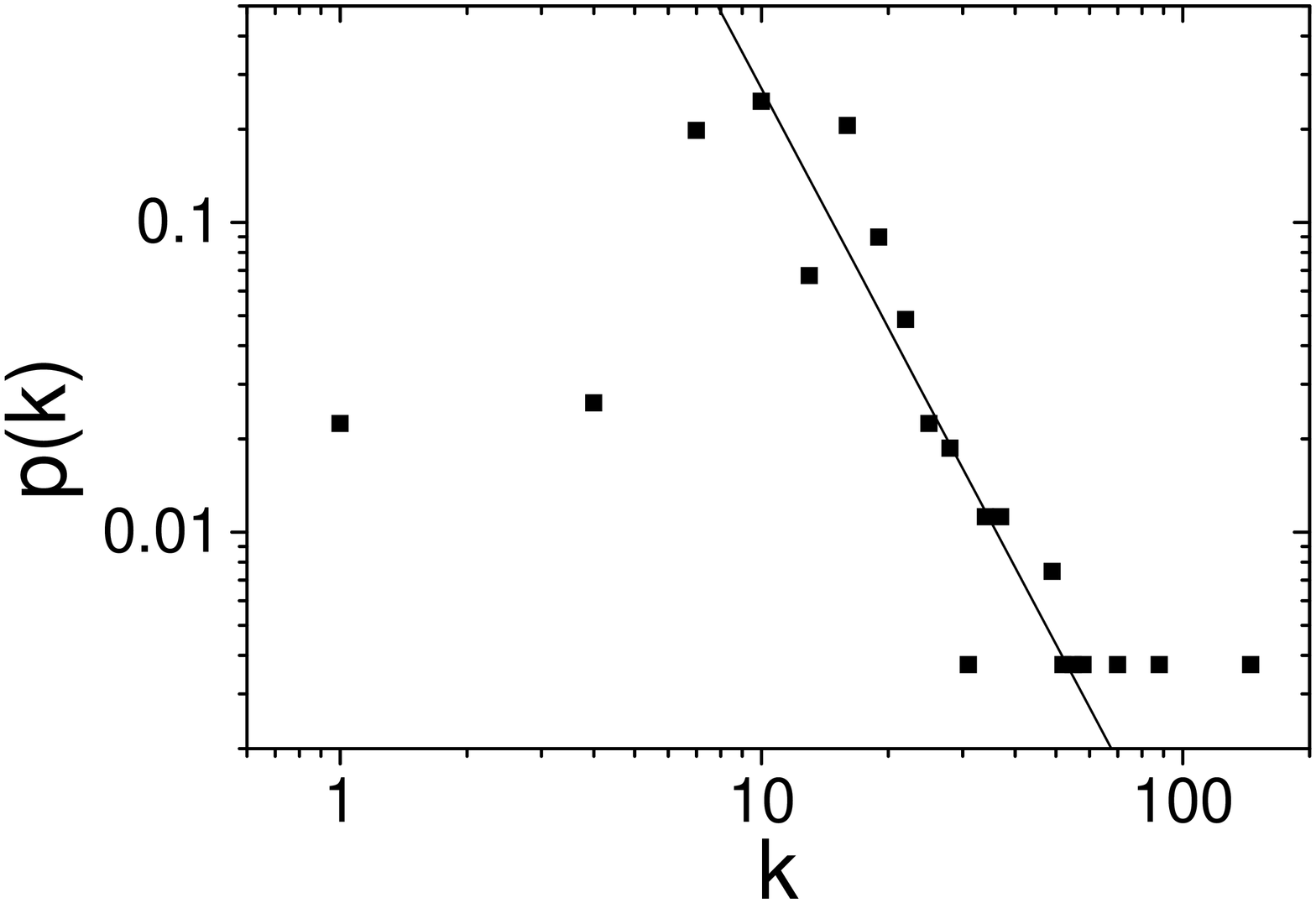}
\end{center}
\caption{Degree distributions $p(k)$ for a very large and the
smallest text.}
\end{figure}

The dynamical analysis also reveals very interesting results. In
Figure 3(a), points indicate the evolution of C for 15 large
texts, as function of the network sizes, showing that they
converge to a value around 0.75. We have identified the data of
two specific texts (in color) to emphasize such phenomenon. This
peculiar behavior of $C$ is also exhibited when we plot its value
as function of the number of sentences $s$ in the different texts,
as shown in Figure 3(b). A functional dependence between $C$ and
$s$ can be written as $C \approx 1-0.08 log_{10}s $.

The investigation of the cluster structure of the network during
its growth has shown that texts evolve mostly in the form of a
very large single cluster, which is formed since the very
beginning of the network evolution. New words are likely to adhere
to it rather than starting other significant clusters, what leads
to the presence of a single giant cluster for the entire text.
Another aspect we have analyzed was the evolution of the indices
in the network growth process with respect to the same text in
different languages. As an example we compare, in Figure 4, the
evolution of $C$ and $\ell$ for the network associated to James
Joyce's Ulysses, in the original English version and in its
translation to Portuguese.

A partial discussion of the results we have obtained to this point
indicates that the networks we analyzed have highly sparse
connectivity, small $D$ and $\ell$, but high $C$, what constitute
evidences of a small-word network scenario. Besides, since $p(k)$
decay according to power laws, we conclude they also behave as
scale-free networks (see values in Table I). It is important to
emphasize that the filtering treatment does not modify the network
general behavior. In order to verify this fact, we have also
performed the same measurements for the networks obtained from the
original texts, without any kind of treatment. We have verified
that, notwithstanding to the fact that we obtain different values
for the same indices (see Table I), the very frequent presence of
articles, prepositions and other words does not alter the
small-world and scale-free character of the text networks! The
small-world behavior is characterized by a great compactness and
by the presence of some vertices with high local clustering. Such
aspect in these networks means that they have words that are often
used along the text. That fact explains the small values of $D$
and $l$. The scale-free behavior results from the presence of
nodes with high connectivity degree in a much greater amount in
comparison to that of a random graph. This aspect is obviously
expected in a written text, due to the presence of an organization
principle which controls the construction of the text. The most
connected words are associated with recurrent concepts around
which the text is constructed.

\begin{figure}
\begin{center}
\includegraphics*[width=4cm,angle=270]{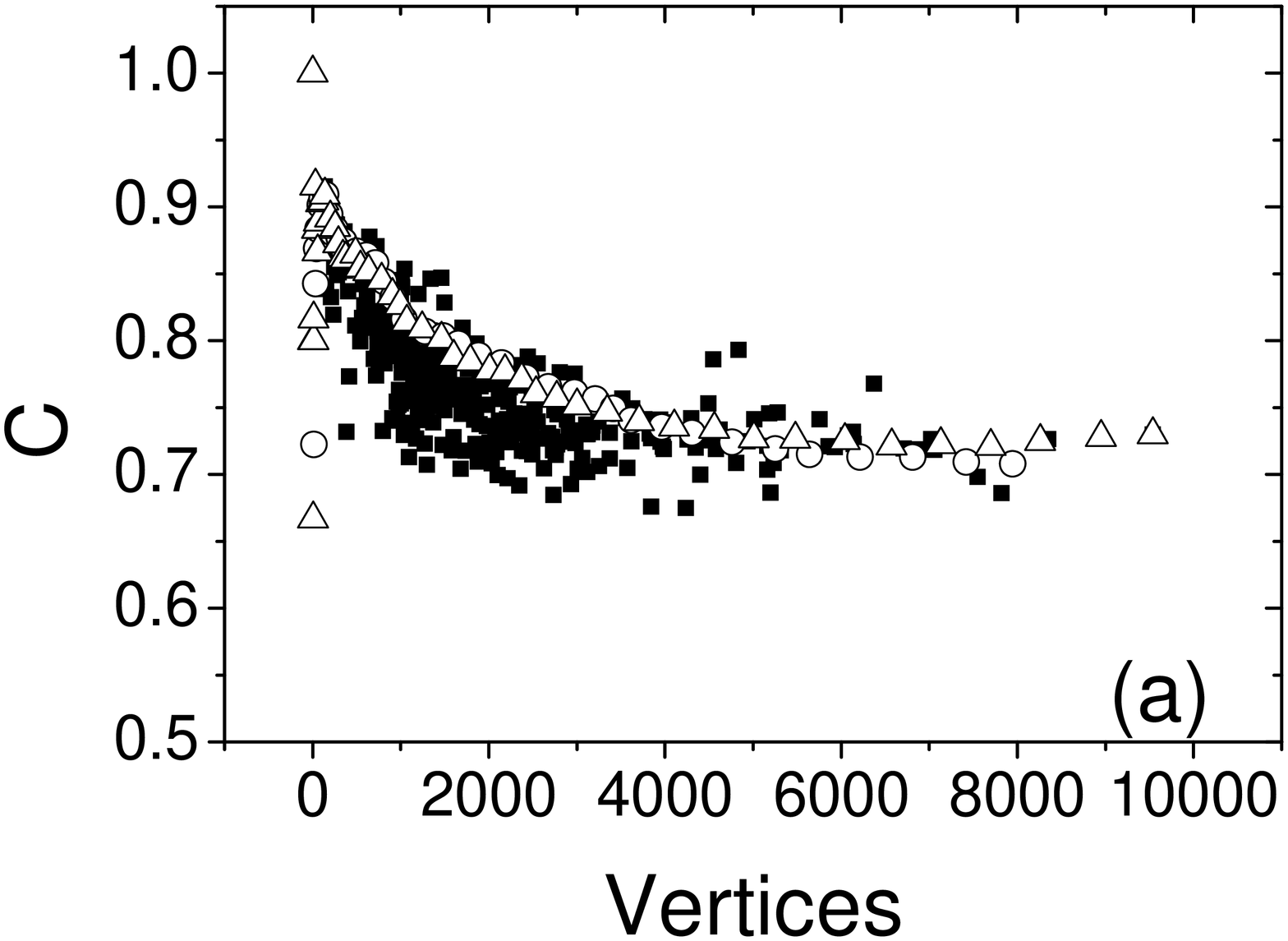}
\includegraphics*[width=4cm,angle=270]{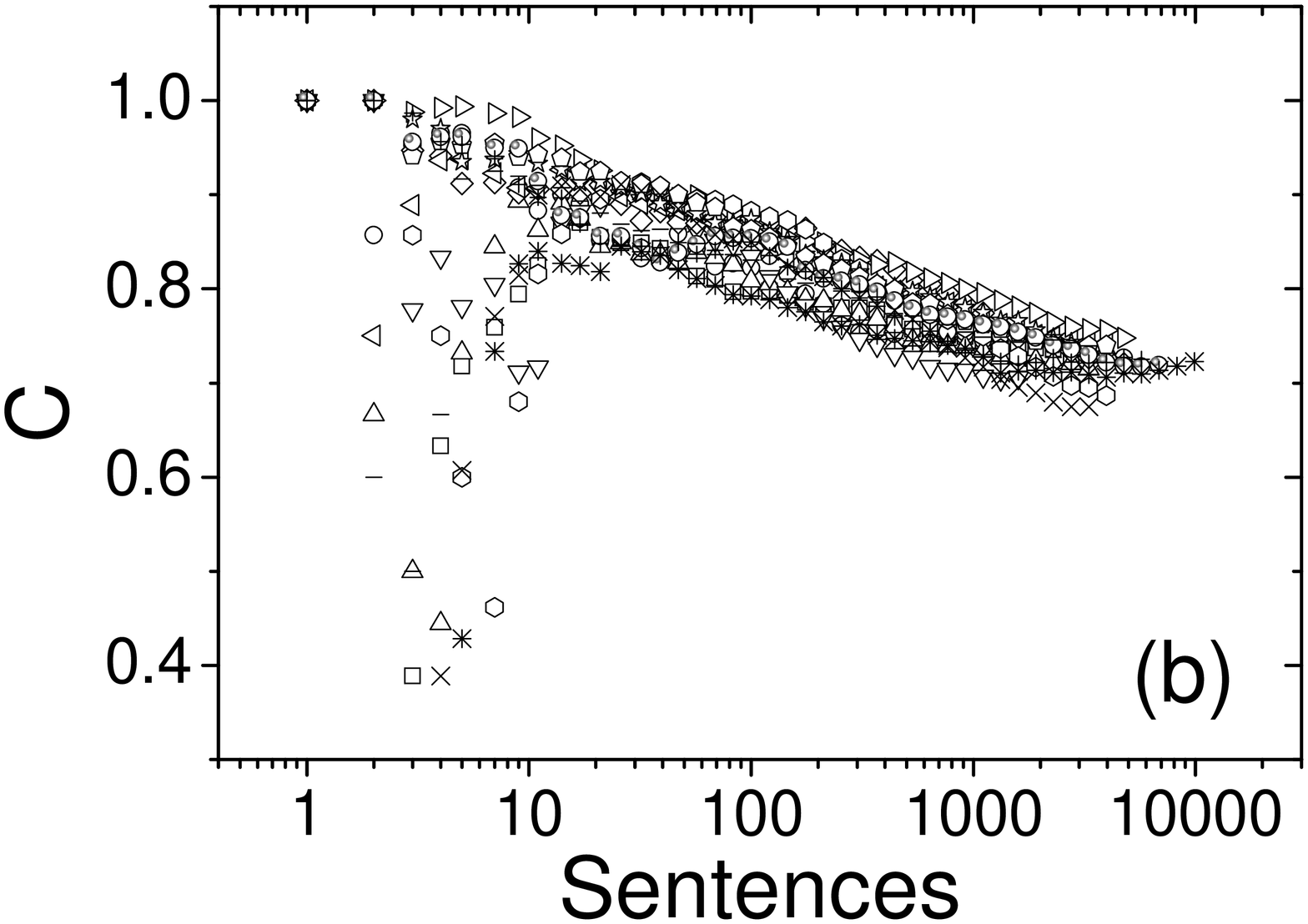}
\end{center}
\caption{(a) Behavior of C as function of the number of vertices
compared with two texts. (b) The same index as function of the
number of sentences, for 15 texts.}
\end{figure}

\begin{figure}
\begin{center}
\includegraphics*[width=4cm,angle=270]{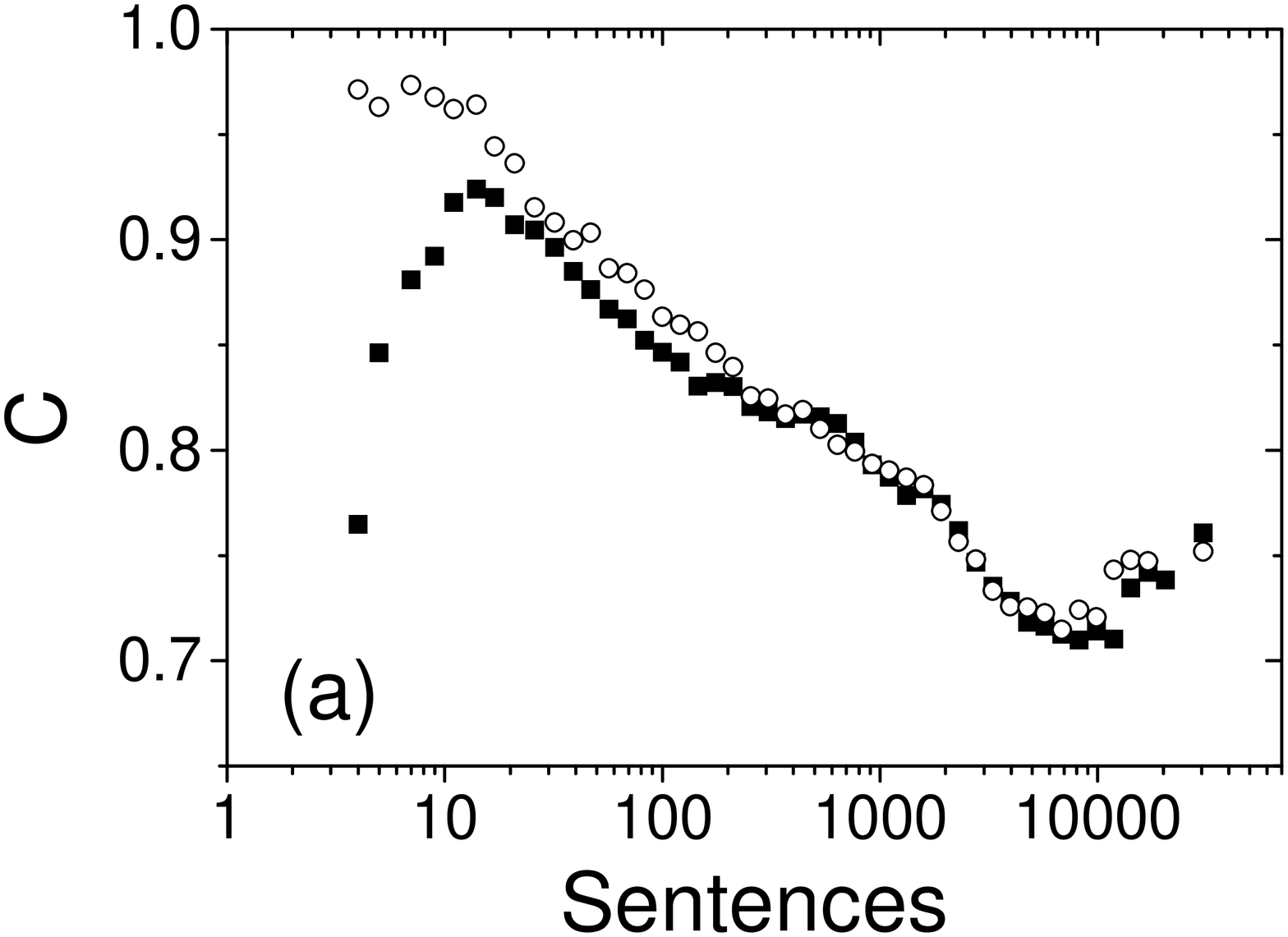}
\includegraphics*[width=4cm,angle=270]{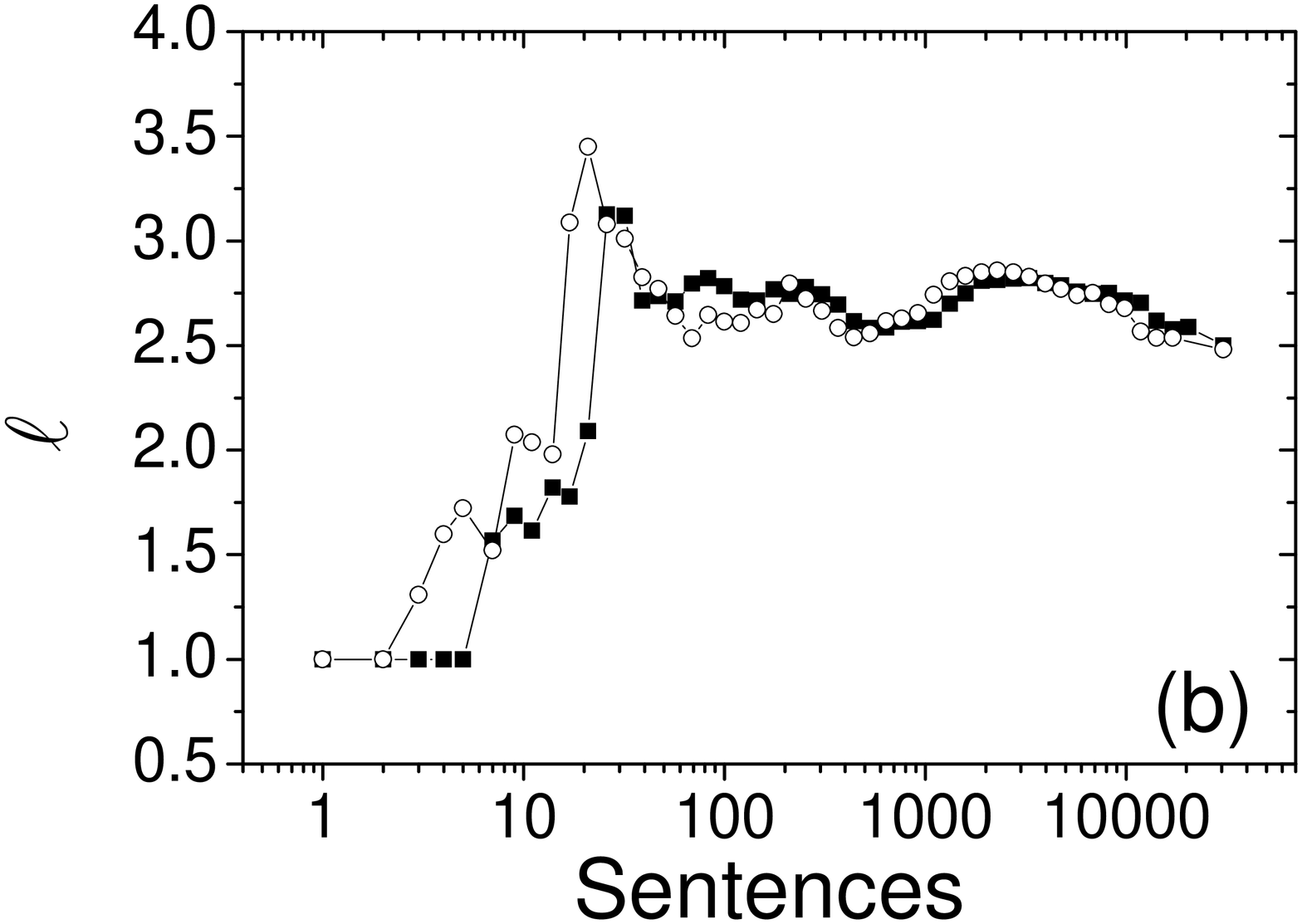}
\end{center}
\caption{Clustering coefficient and average shortest path length
evolution, according to the number of sentences in James Joyce's
Ulysses, in the original English version and in a Portuguese
translation.}
\end{figure}

\section{Shuffling procedures}

There are several agents that contribute to determine the measured
indices in the analyzed networks. For example, since each sentence
is added to the network as a complete subgraph, their lengths may
interfere in the clustering coefficient. In the same way, the own
structure of the sentences and the frequency of the words along
the whole text affects C and other indices as well. In order to
evaluate the role of these main agents in the determination of the
indices, we have re-analyzed the filtered texts after submitting
them to four shuffling processes, which are so characterized:
$(SA)$ In this process, the beginning and the end of sentences are
kept fixed, however the position of the words along the whole text
is randomly changed. This procedure breaks completely the
structure of concepts in the sentences, but their lengths and the
frequency of the words in the whole text are kept unchanged.
$(SB)$ Here, the original sequence of the words along the text
remains unchanged, but all sentences are forced to have the same
average length, as obtained from the analysis of the unperturbed
text. $(SC)$ As in $SA$, the beginning and the end of sentences
are kept fixed, and words are randomly chosen from the same
vocabulary of the text. However, the words frequency is changed to
a white distribution, so that all the words have the same choice
probability. $(SD)$ Here, we randomly distribute the same number
of links as in the network of the filtered text among the nodes,
i.e., obtaining an Erd\"{o}s-Renyi network.

\begin{figure}
\begin{center}
\includegraphics*[width=5cm,angle=270]{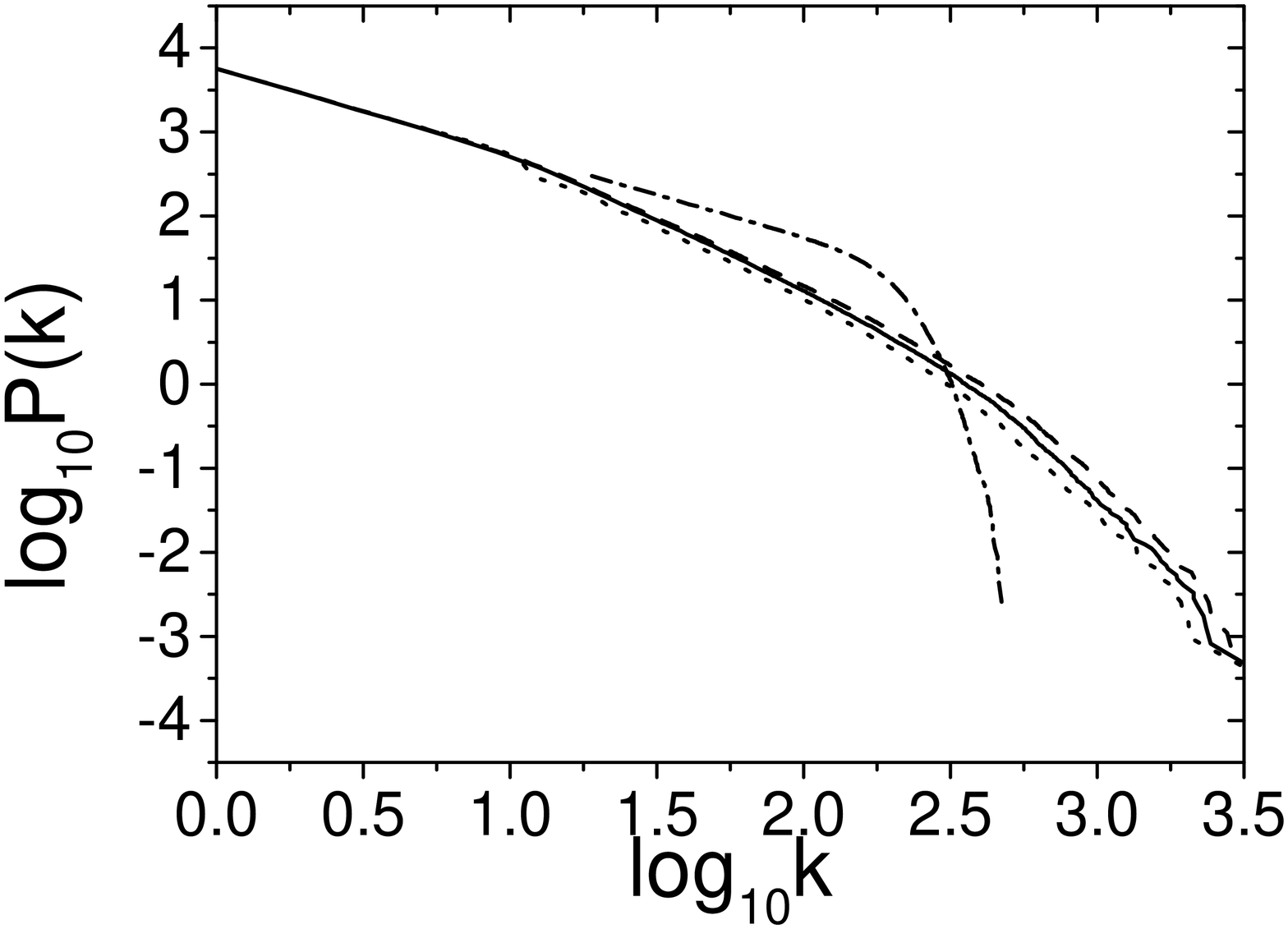}
\end{center}
\caption{Cumulative degree distributions $P(k)$ obtained with one
single text, for the filtered version (solid), and the shuffle
procedures $SA$ (dashed), $SB$ (dotted), and $SC$ (dash-dotted).}
\end{figure}

A summary of results is also included in the Table I. They show
that the networks are affected in different ways but, as expected,
for all but the $SD$ procedure, they are very far from random
networks. The indices for $SA$ and $SB$ are not significantly
altered. In the first case it shows that, without changing the
structural aspects of sentence sizes and word frequencies, the
network is not essentially affected. In the second one, breaking
sentence sizes but keeping words in their original places is not
sufficient to alter, in a meaningful way, the network structure.
These traits are corroborated by Figure 5, where we show the 4
cumulative degree distribution

\begin{equation}
\label{eq1}P(k)=\frac{1}{k}\int _k ^{\infty}p(k')dk',
\end{equation}

for a given text and three shuffled versions of it. We realize
that the scale-free character, expressed by a linear decay in the
$k$ interval [10,300] for $P(k)$ remains almost unchanged. These
results remind Zipf's universal result about distribution
frequency of words in texts since, when Zipf's law is not
followed, e.g., $SC$ and $SD$, the network structure is broken.
Indeed, not only the power law for $P(k)$ is lost, but also the
indices in Table I have been altered. Nevertheless we emphasize
the fact that, despite the change in word frequency caused by $SC$
sets up a deep modification in the text, its value for $C$ is much
higher than that one for $SD$. This indicates that the unchanged
distribution of sequence sizes still keeps the $SC$ network far
away from the Erd\"{o}s-Renyi scenario.

The results of the dynamical analysis call our attention to
general aspects of network growth. Barab\'{a}si and
Albert\cite{r10} observed that the scale-free behavior may be
explained by a special kind of growth process, known as
preferential attachment. They proposed a model that captures such
behavior: vertices are added to the network, systematically, by
connecting them to some just existing vertices, which are selected
in accordance with a probability distribution that depends on
their degrees. However, the values obtained for $C$ are much lower
than those in small-world networks. On the other hand, the Watts
and Strogatz small-world, obtained by randomly rewiring a regular
network, does not reproduce the scale-free feature\cite{r9}
Therefore, we may suggest that the growth process we employ here
is essential to capture both quoted behaviors. It is characterized
by the addition of a new sentence, i.e., a complete subgraph or
clique in each step. Due to the frequency distribution of words
along the text, the attachment of these complete subgraphs is
still preferential, but the new vertices are highly clustered,
what contributes to the coexistence of both small-world and
scale-free network scenario. Thus, the model of preferential
attachment by complete subgraphs according to a pre-established
frequency of vertices seems to be more suitable to describe the
concomitant emergence of these behaviors.

\begin{table}
\begin{center}
\begin{tabular}{|c|c|c|c|}\hline
Text & $L$ & $\ell$ &  $C$  \\\hline
Filtered & 5$\pm$ 1 & 2.3$\pm$ 0.2 & 0.77$\pm$ 0.05 \\
$SA$ & 4$\pm$ 1 & 2.2$\pm$ 0.2 & 0.77$\pm$ 0.04 \\
$SB$ & 4$\pm$ 1 & 2.4$\pm$ 0.3 & 0.74$\pm$ 0.05 \\
$SC$ & 4$\pm$ 1 & 2.2$\pm$ 0.3 & 0.40$\pm$ 0.20\\
$SD$ & 3$\pm$ 1 & 2.2$\pm$ 0.3 & 0.05$\pm$ 0.06 \\
Original & 4$\pm$ 1 & 2.0$\pm$ 0.1 & 0.82$\pm$ 0.03\\ \hline
\end{tabular}
\end{center}
\caption{Average values for the indices. Filtered texts were used
as standard subjects, and shuffling operations were carried out on
networks generated by them. Original means text without filtering,
used just for the purpose of comparison. }
\end{table}

To conclude, we would like to stress that our major purpose, to
analyze network structure among concepts expressed by significant
words in written texts, has been successfully addressed. The
obtained indices point to the presence of both small-world and
scale-free features. Moreover, we also characterized important
aspects observed as the text grows, and studied the influence of
distinct shuffling procedures of words and sentence size
distribution. As expressed in the first paragraphs, our results
may be of significance to shed light into the deeper problem of
how the human mind works. We have briefly mentioned some efforts
concerning the construction and analyzes of networks that
represent the relations between neurons in brain cortex\cite{r17},
such networks presenting both small-world and scale-free behavior.

The relation mind-brain is a subject of great controversy, but we
wonder whether the universal character of the network behavior we
have found in our work does not point out to some patterns present
in the human mind. Indeed, due to the intense intellectual
activity required to produce texts, it is natural to expect that
the results may help understanding the way our mind works.

Patterns that might control the way we store concepts in our minds
and could reveal aspects of the structure of our brains. It is
also worthy to remember that similar patterns to those we obtained
here have been observed in networks of semantically related words.
Thus, it is natural to conjecture that the results we found do
reveal the existence of links between all these research areas.
Certainly, there is much to be explored in such text networks, as
this is a vast field and the possibility of a joint effort from
several different areas of investigation is open.

Acknowledgements: The authors would like to thank Fernanda Regebe,
Gesiane Teixeira and Charles Santana for helpful discussions and
assistance by code programming.


\begin{thebibliography}{00}

\bibitem{r1} {D. J. Watts, {\it Small Worlds: The
dynamics of networks between order and randomness}, (Princeton
University Press, Princeton 1999).}
\bibitem{r2} {M. Buchanan,{\it Nexus: small world and the groundbreaking science of networks},
(W. W. Norton \& Company, Inc., New York, 2002).}
\bibitem{r3} {A.-L. Barabasi, \textit{Linked: How everything
is connected to everything else and what it means}, (Plume
Printing, New York 2003).}
\bibitem{r4} {S.N. Dorogovtsev, and J.F.F. Mendes, \textit{Evolution of
Networks}, (Oxford University Press, Oxford 2003).}
\bibitem{r5} {R. Albert, and  A.-L. Barabasi, Rev. Mod. Phys. \textbf{74},
47 (2002).}
\bibitem{r6} {J. Camacho, R. Guimera, and  L.A.N. Amaral, Phys. Rev. Lett.
\textbf{88}, 228102 (2002).}
\bibitem{r7} {A.-L. Barabasi, R. Albert, and H. Jeong,  Nature \textbf{401},
130 (1999).
\bibitem{r8} {J. Guare, \textit{Six Degrees of Separations: A Play},
(Vintage, New York 1990).}
\bibitem{r9} {D. J. Watts, and S. D. Strogatz, Nature \textbf{393},
440 (1998).}
\bibitem{r10} {A.-L. Barabasi, and R. Albert, Science \textbf{286}, 509
(1999).}
\bibitem{r11} {A.E. Motter, A.P.S. de Moura, Y.C. Lai, and P. Dasgupta,
Phys. Rev. E \textbf{65}, 065102 (2002).}
\bibitem{r12} {L. F. Costa, Intl. J. Mod. Phys. C \textbf{15}, 371 (2004).}
\bibitem{r13} {R.V. Sole, Nature \textbf{434}, 289 (2005).}
\bibitem{r14} {R.F.i. Cancho, and R.V. Sole, Proc. R. Soc. London, Ser. B \textbf{268}, 2261 (2001).}
\bibitem{r15} {R.F.i. Cancho, R.V. Sole, and R. Kohler, Phys. Rev. E 69, 051915 (2004).}
\bibitem{r16} {S.N. Dorogovtsev, and J.F.F. Mendes, Proc. R. Soc. London, Ser. B \textbf{268},
2603 (2001).}
\bibitem{r17} {V. Eguiluz, G. Cecchi, D.R. Chialvo, M. Baliki, and A.V. Apkarian, Phys. Rev. Lett. \textbf{92}, 018102
(2005).}
\bibitem{r18} {G.K. Zipf, \textit{Human behaviour and the principle of
least effort. An introduction to human ecology}, (Hafner, New
York, 1972).}
\bibitem{r19} {Unitex is a multilingual corpus
processing system, based on automata-oriented technology developed
by the Computational Linguistics Group of the Institut
d'eletronique et d'informatique Gaspard-Monge (IGM-France) -
http://infolingu.univ-mlv.fr/english/.}
\bibitem{roda1} {Literary texts have been mostly obtained from
Gutenberg Project website at www.gutenberg.org}
\bibitem{r20} {Freud, S., \textit{Zur Auffassung der Aphasien}, (Leipzig,
Deuticke, 1891). Available in English as \textit{On Aphasia: A
Critical Study} (Translated by E. Stengel, International
Universities Press, 1953).}}
\end{thebibliography}
\end{document}